\documentclass[12pt,preprint]{aastex}

\slugcomment{Submitted to the Astrophysical Journal}

\shorttitle{The expanding bipolar shell of the helium nova V445 Puppis}
\shortauthors{Woudt et al.}

\begin{document}

\title{The expanding bipolar shell of the helium nova V445 Puppis}

\author{P. A. Woudt}
\affil{Department of Astronomy, University of Cape Town, Private Bag X3,
    Rondebosch 7701, South Africa}
\email{Patrick.Woudt@uct.ac.za}

\author{D. Steeghs\altaffilmark{1}}
\affil{Department of Physics, University of Warwick, Coventry, CV4 7AL, 
United Kingdom}

\author{M. Karovska}
\affil{Harvard-Smithsonian Center for Astrophysics, 60 Garden Street, 
Cambridge MA 02138-1516, USA}

\author{B. Warner\altaffilmark{2}}
\affil{Department of Astronomy, University of Cape Town, Private Bag X3,
    Rondebosch 7701, South Africa}

\author{P. J. Groot and G. Nelemans}
\affil{Department of Astrophysics, Radboud University Nijmegen, PO Box 9010, 
6500 GL, Nijmegen, the Netherlands}

\author{G. H. A. Roelofs}
\affil{Harvard-Smithsonian Center for Astrophysics, 60 Garden Street, 
Cambridge MA 02138-1516, USA}

\author{T. R. Marsh}
\affil{Department of Physics, University of Warwick, Coventry, CV4 7AL, 
United Kingdom}

\author{T. Nagayama}
\affil{Department of Astronomy, Kyoto University, Kitashirakawa-Oiwake-cho, 
Sakyo-ku, Kyoto 606-8502, Japan}

\author{D. P. Smits}
\affil{Department of Mathematical Sciences, University of South Africa, UNISA, Pretoria 0003,
South Africa}

\and

\author{T. O'Brien}
\affil{University of Manchester, Jodrell Bank Observatory, Macclesfield, 
Cheshire SK11 9DL, United Kingdom}

\altaffiltext{1}{Center for Astrophysics,
    60 Garden Street, Cambridge, MA 02138-1516}

\altaffiltext{2}{School of Physics and Astronomy, Southampton University,
    Highfield, Southampton SO17 1BJ, United Kingdom}

\begin{abstract}
From multi-epoch adaptive optics imaging and integral field unit 
spectroscopy we report the discovery
of an expanding and narrowly confined bipolar shell surrounding 
the helium nova V445 Puppis (Nova Puppis 2000). An equatorial dust
disc obscures the nova remnant, and the outflow is characterised
by a large polar outflow velocity of 6720 $\pm$ 650 km s$^{-1}$ and 
knots moving at even larger velocities of 8450 $\pm$ 570 km s$^{-1}$. 
We derive an expansion parallax distance of 8.2 $\pm$ 0.5 kpc and 
deduce a pre-outburst luminosity of the underlying binary of 
$\log L/L_{\odot} = 4.34 \pm 0.36$. The derived luminosity suggests that
V445 Puppis probably contains a massive white dwarf accreting at high rate from 
a helium star companion making it part of a population of binary stars that
potentially lead to supernova Ia explosions due to accumulation of helium-rich 
material on the surface of a massive white dwarf.

\end{abstract}


\section{Introduction}

Accretion onto compact objects can lead to a range of explosive phenomena since 
their accreted layers can reach densities and temperatures high enough to initiate 
nuclear burning.
Helium novae are expected to occur during periods of high mass-accretion rates 
($\dot{M} \sim 10^{-9} - 10^{-6}$ M$_{\odot}$ yr$^{-1}$) of He-rich material 
through unstable helium burning via helium shell flashes \citep{iben91}. 
In such events, typically $\sim 10^{-4}- 10^{-2}$ M$_{\odot}$ of fuel is ignited 
on the surface of the accretor \citep{kato99,yoon04,bild07}.
Models of the binaries that  experience such helium novae generally consist of a 0.6 -- 
0.8 M$_{\odot}$ CO white dwarf (the accretor) with a lower-mass H-deficient companion (the donor). 
The donor can be either a non-degenerate helium star \citep[e.g.,][]{yoon04} or a semi-/fully 
degenerate star. The latter are usually referred to as AM CVn
systems, representing a class of ultra-compact helium-transferring binaries 
\citep{nele04,bild07,roel07}. 

Some of these compact binaries may lead to sufficient mass accumulation onto the primary 
to be viable supernova Ia progenitors after successive He 
novae \citep{iben94,kato89,yoon03,tutu07,wang09}. Accreting massive white dwarfs in close 
binary systems are the favoured progenitors of supernova Ia explosions \citep{branch07,par07}. 
Possible binary  channels are typically divided into single-degenerate scenarios involving 
hydrogen-rich companions versus double-degenerate progenitors. The latter include the white 
dwarf plus He-star and double white dwarf pathways.
To date, only about a dozen promising single-degenerate type Ia progenitors 
have been identified \citep[see, e.g.,][]{par07}. U~Scorpii and 
RS Ophiuchi represent the two different classes of hydrogen-rich companion 
stars in the supernova Ia progenitors: main-sequence or slightly evolved stars, and 
red giants, respectively \citep{li97}.
Ongoing supernova searches are, however, revealing a growing diversity of 
type Ia explosion events, which challenge current formation scenarios. It could
be that, despite their low predicted frequency, the helium star donor channel to 
type Ia explosion is successful in explaining some of the diversity 
observed, such as the population of young Ia progenitors \citep{manu06,wang09}.

\object{V445 Puppis} (Nova Puppis 2000) is the first, and so far 
only, helium nova detected \citep{ashok03,kato03}. The nova outburst of V445
Puppis was first noticed on 23 November 2000 \citep{kato01}, and the optical outburst 
characteristics are those of a slow nova. Optical
\citep{wagn01,iijim08}, 
near-infrared \citep{ashok03} and mid-infrared \citep{lynch01} spectra of V445
Puppis obtained during outburst immediately revealed its most peculiar and defining 
characteristic: the complete absence of hydrogen in the ejecta. 

The outburst light curve has been modelled \citep{kato03,kato08} by free-free 
emission and an optically thick wind, following a helium shell flash 
on the surface of an accreting massive white dwarf. A lower limit on 
the mass of the accreting white dwarf of $M$ $\ge$ 1.35 M$_{\odot}$ is 
inferred by \citet{kato08}, making V445 Puppis a possible progenitor of 
a supernova Ia from a helium-rich donor channel. In these models, the white dwarf is expected to 
grow in mass as the mass ejected through repeated helium nova outbursts is less than 
the accreted mass \citep{kato99}. 

V445 Puppis provides the first empirical benchmark for a helium-dominated outburst 
on the surface of a white dwarf. 
In Section 2 we present new observations of V445 Puppis taken 5 -- 7 
years after outburst, and determine the distinct evolution of the resolved 
nova shell from a spatio-kinematic analysis (Section 3). This leads to a robust
distance determination to V445 Puppis. 
In Section 4, we discuss the implications of the distance 
derived here on the nature of the underlying binary.

\section{Imaging and Spectroscopic Observations}

We acquired high angular-resolution images of V445 Puppis 
in the near-infrared $K_s$ band using the NAOS/CONICA adaptive optics (AO) 
system \citep{rous03,lenz03} on the Very Large Telescope (VLT). 
For these observations, the S27 camera was used which has a scale of
27.15 milli-arcsec per pixel and a total field of $28'' \times 28''$. 
The images were taken on four epochs spread over a two-year period; details
of the VLT observations are given in Table \ref{woudttab1}. 
The nearby ($24''$) and relatively bright 
($V \sim 15.5$ mag) star GSC2.3 S3EQ038054 \citep{lask08} was selected as 
the AO reference star. The achieved
FWHM on target after AO corrections ranges from $0.09''$ to $0.14''$.
Separate observations of a pair of close stars -- also separated by 
$24''$ and of similar brightness to GSC2.3 S3EQ038054 -- were 
obtained for deconvolution purposes; the selected PSF star is GSC2.3 
S3EQ038567 \citep{lask08}.  We deconvolved the images using the standard 
Richardson-Lucy technique \citep{rich72,luc74,karo97}. It took from 5 to 20 
iterations for the deconvolution to converge to a stable result.

In addition, integral field unit (IFU) spectroscopy was obtained with the 
IMACS spectrograph on the 6.5-m Magellan Baade telescope 
on 4 January 2006. In the f/4 camera mode that we employed, the IMACS IFU mode 
provides two fiber bundles covering $5''\times4.15''$ on the sky sampled with 0.2$''$ 
per fiber element \citep{schmo04}. Spectral coverage was 4465 -- 7634 \AA~sampled at 
0.38 \AA/pixel across four 2k$\times$4k CCD detectors using the 600 grating. We acquired seven 
30-min exposures with V445 Puppis centered in one of the fiber bundles under $0.5''$ seeing 
conditions with clear skies. The second fiber bundle was used to measure simultaneously
the sky $60''$ away from our target. In order to improve the spectral extraction and 
remove cosmic rays, the seven exposures were first median-combined in 2D. Fiber spectra 
were then extracted using a normal extraction procedure covering relevant target and 
sky fibers. Target and sky spectra were wavelength calibrated using He-Ne-Ar lamp 
exposures, and then differenced to produce sky-subtracted 1D fiber spectra sampling 
the target bundle. On three occasions $V$ band imaging was secured with IMACS. 
These images were calibrated using Landolt standards \citep{land92} and the derived magnitudes 
are plotted in Fig.~\ref{woudtfig1}.

Supplementary optical (3600 -- 9000 {\AA}) and near-infrared (15000 -- 25000 {\AA})
long-slit spectroscopy was obtained on 17 December 2003 and 6 October 2005.
Details of the former are presented in \citet{wou05}. The near-infrared spectroscopy
was obtained in service mode on the 3.5-m New Technology Telescope using the SOFI 
spectrograph. The spectral coverage includes the $H$ and $K_S$ photometric bands, sampled
at 10.2 {\AA}/pixel.

We frequently monitored V445 Puppis at near-infrared wavelengths, 
from 2002 March to the present, using the Infrared Survey Facility and the
SIRIUS camera \citep{naga03} at the Sutherland station of the South African
Astronomical Observatory; the latter allows simultaneous $J$, $H$ and
$K_s$ imaging over a $7.7 \times 7.7$ arcmin field of view. The IRSF aperture-corrected photometry has
been calibrated with photometry from the 2MASS Point Source Catalog \citep{skrut06} using
stars in the field of view. 2MASS magnitudes of these stars have been converted to 
the IRSF photometric system to determine proper zero points, after which the 
calibrated IRSF magnitudes of V445 Puppis have been converted back to the 2MASS system. 
The magnitudes of V445 Puppis in the 2MASS photometric system are given in Table~\ref{woudttab2}.

\section{Results}

\subsection{Brightness evolution}

\begin{figure}
\epsscale{0.8}
\plotone{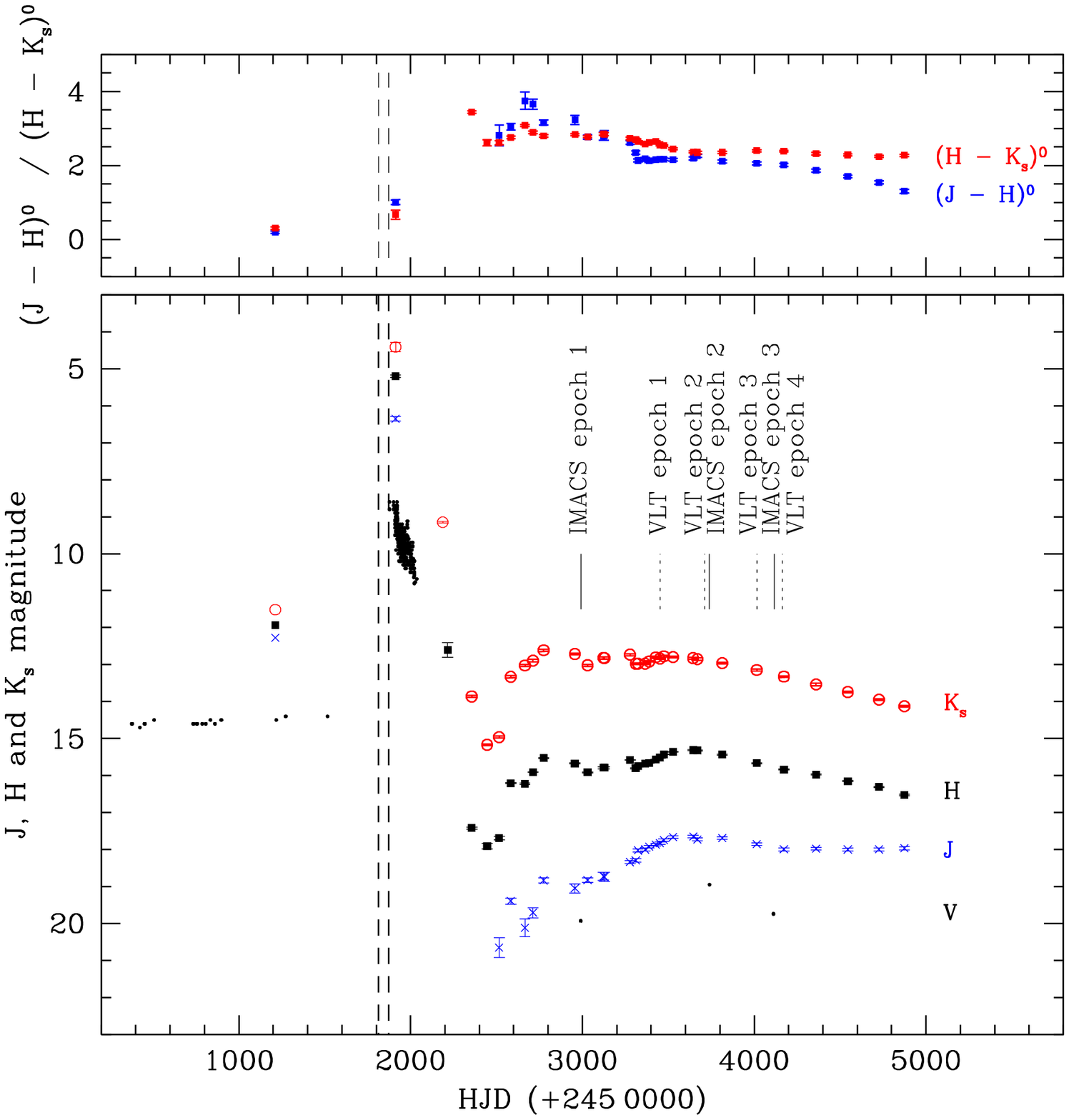}
\caption{The optical and near-infrared light curve of V445 Puppis before, during and 
after outburst (lower panel). The exact date of outburst is uncertain, but constrained by the two vertical
dashed lines. The times of VLT observations (Fig.~\ref{woudtfig2}) are indicated by the vertical dotted lines, times of
Magellan/IMACS observations are indicated by the vertical solid lines.
The top panel shows the near-infrared color evolution of V445 Puppis which is corrected for Galactic
foreground extinction.}
\label{woudtfig1}
\end{figure}

The long-term brightness evolution of V445 Puppis at optical and near-infrared 
wavelengths is shown in the lower panel of Fig.~\ref{woudtfig1}. It shows V445 Puppis before outburst in the 
$V$ band and the 2MASS $J$, $H$ and $K_s$ single epoch \citep{skrut06}, during outburst in optical
\footnote{All optical pre-outburst and outburst observations are from VSNET, see \citet{ashok03} for details.} and
near-infrared \citep{ashok03}, and post-outburst where 
the initial $H$ and $K_s$ observation after decline is from \citet{ashok03} and 
the remainder of the post-outburst light curve is derived from our 
IRSF/SIRIUS $J$, $H$, $K_s$ and Magellan/IMACS $V$ band observations. 
The exact date of outburst is uncertain, but is constrained by 
a lower detection limit ($V < 12$ mag) of V445 Puppis less than two months prior to its discovery in outburst,
see the discussion in \citet{ashok03}. The outburst has occured within the two vertical dashed lines 
in Fig.~\ref{woudtfig1}. The four epochs of high-angular resolution imaging with NAOS/CONICA are
indicated by the vertical dotted lines, the three epochs of IMACS observations are marked by solid 
vertical bars.

More than 8 years after outburst, V445 Puppis is still 
highly reddened and $\sim$ 6 magnitudes below its pre-outburst brightness as measured in the $J$ band and 5.2 magnitude below
its pre-outburst brightness in the $V$ band; at this phase, optical light is dominated by line emission with a 
negligible continuum contribution. In the last
2 -- 3 years, V445 Puppis has steadily declined in brightness in the $H$ and $K_s$ bands, but stayed constant
in the $J$ band. Where the $J$ band flux has almost 
no continuum component and is dominated by the 1.0830 $\mu$m He\,I recombination 
line emission, the $H$ and $K_s$ bands contain a thermal continuum component of $T = 250$ K \citep{lynch04}. The
latter band also contains the 2.0581 $\mu$m He\,I recombination line which has an equivalent width of $\sim -180$~{\AA}
as derived from our SOFI spectrum. 
The upper panel of Fig.~\ref{woudtfig1} shows the $(J-H)^0$ and $(H-K_s)^0$ color
evolution of V445 Puppis, where the photometry has been corrected for the Galactic foreground extinction
of $E(B-V) = 0.51$ mag \citep{iijim08}. We refer to Sect. 4.1 for a more detailed discussion on 
the Galactic foreground extinction.
The fading in the $H$ and $K_s$ band -- at fairly constant $(H-K_s)^0$ color -- could indicate a cooling of the
thermal continuum component over the last 2 -- 3 years. {\it Spitzer} observations of V445 Puppis
taken during this phase (PI: Banerjee) will be able to constrain the temperature of the cool thermal component.

\subsection{The expanding nova shell}

\begin{figure*}
\epsscale{1.0}
\plotone{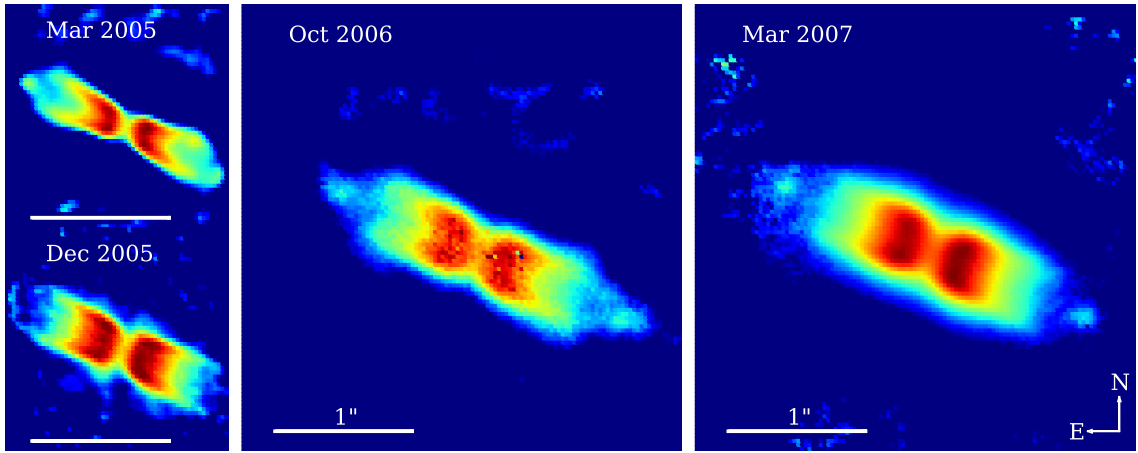}
\caption{The evolving nova shell of V445 Puppis over a period of 2 years,
obtained on four different epochs with NAOS/CONICA on the Very Large Telescope. 
All images are plotted on the same scale in the conventional North-East configuration
with a sampling of 27.15 mas/pixel. The horizontal bar denotes a length of 1$''$.}
\label{woudtfig2}
\end{figure*}

\begin{figure}
\epsscale{0.5}
\plotone{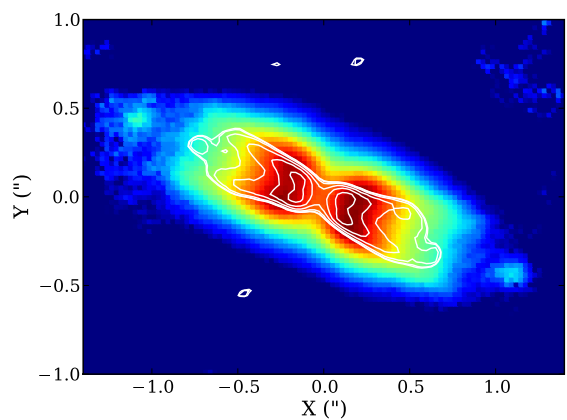}
\caption{Two epochs of NAOS/CONICA imaging (March 2005 and March 2007) superimposed on top of
each other (March 2005: contours, March 2007: false colour), revealing the distinct evolution of the shell.
A spatial scale in arcseconds is provided on the both coordinate axes. North is up and East is to the left.}
\label{woudtfig2b}
\end{figure}

Fig.~\ref{woudtfig2} shows the deconvolved near-infrared $K_s$ (false color) images 
obtained at the four different epochs. The deconvolved field of view of the March
and December 2005 images is $1.66'' \times 1.52''$, whereas the October 2006 
and March 2007 observations are displayed on a wider scale of 
$3.48'' \times 3.48''$.  The incredible detail seen on these small scales -- 
the ejecta only span $2'' - 3''$ on the sky -- is possible thanks to the adaptive optics 
technology on large ground-based telescopes. This allows a clear morphological description 
and dynamical analysis of the evolving nova ejecta. In Fig.~\ref{woudtfig2b} we show the 
shell in March 2005 (in contours) superimposed on the March 2007 observations 
to demonstrate the distinct expansion of the shell.

The images unambiguously show a bipolar shell -- initially with a very narrow
waist -- with lobes on each side (north-east: NE, and south-west: SW) of a 
centrally-obscured region covering the nova remnant.  At all epochs, two knots
are seen at either extreme end of the nova shell.
This shell is unlike any previously observed classical nova shell, 
though strongly reminiscent of bipolar planetary nebulae (PNe), protoplanetary 
nebulae (pPNe) \citep{balick02}, and the symbiotic Mira nebula Hen 2-104
\citep{cor01}. In the case of V445 Puppis though, the large expansion velocity of the shell
is clearly consistent with a nova explosion, rather than a PN outflow. 
The central obscuration appears to be confined 
to a plane nearly perpendicular to the two lobes of the bipolar shell. 
Such an obscuration has also been seen in various pPNe 
\citep[e.g. in Hen\,401,][]{sahai99} where it is interpreted as a thick 
circumstellar dust disk. As the ejecta of V445 Puppis had a dominant carbon 
signature shortly following the outburst, and as the nova remnant has now been
obscured for the last 8 years following outburst, it is likely that this dust 
is largely concentrated in a plane perpendicular to the lobes where it 
continues to obscure the underlying binary. Given the large obscuration post-outburst (see lower panel
Fig.~\ref{woudtfig1}), this dust disk must be a direct consequence of the November 2000 outburst.

\subsection{Spatio-kinematic analysis}

\begin{figure}
\epsscale{1.0}
\plottwo{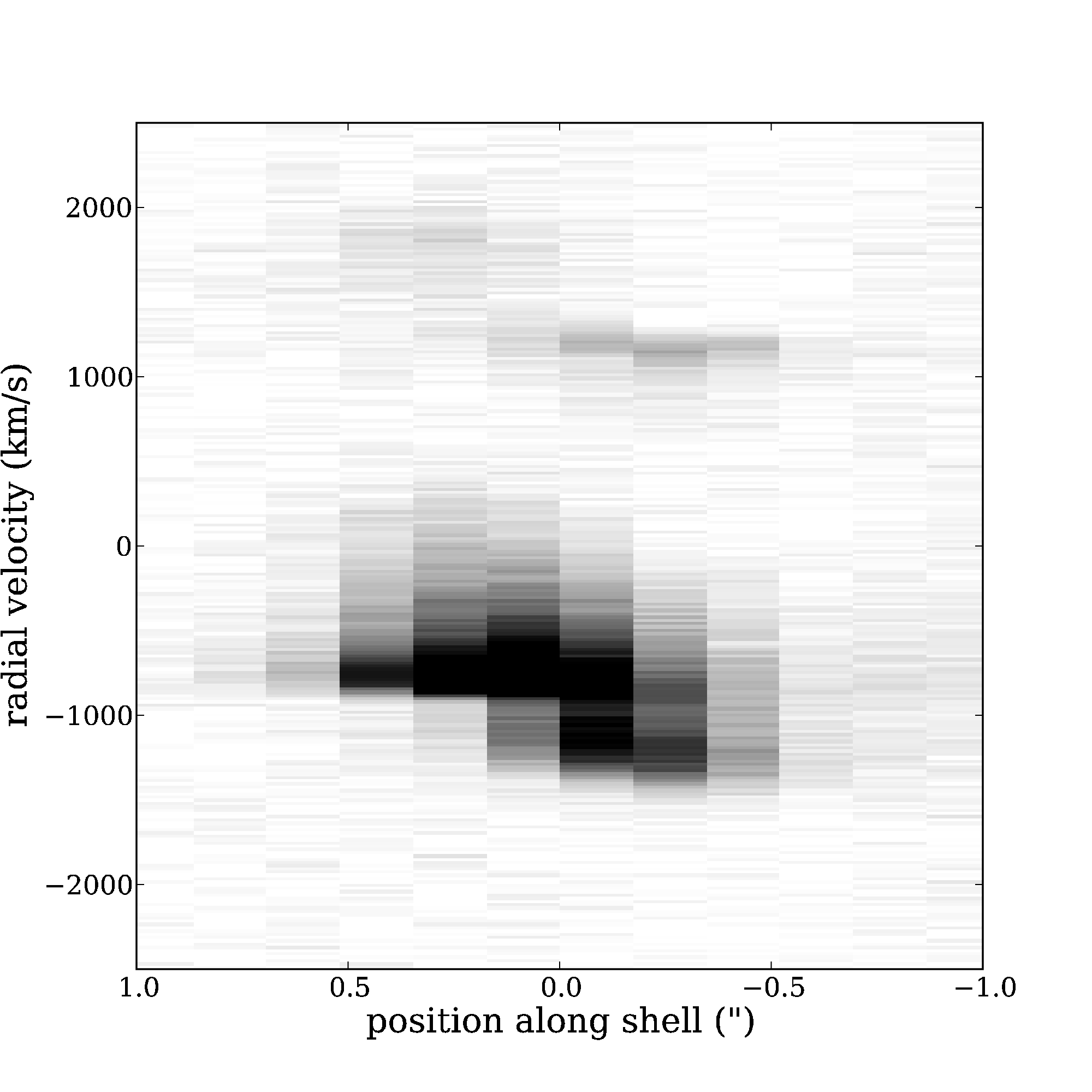}{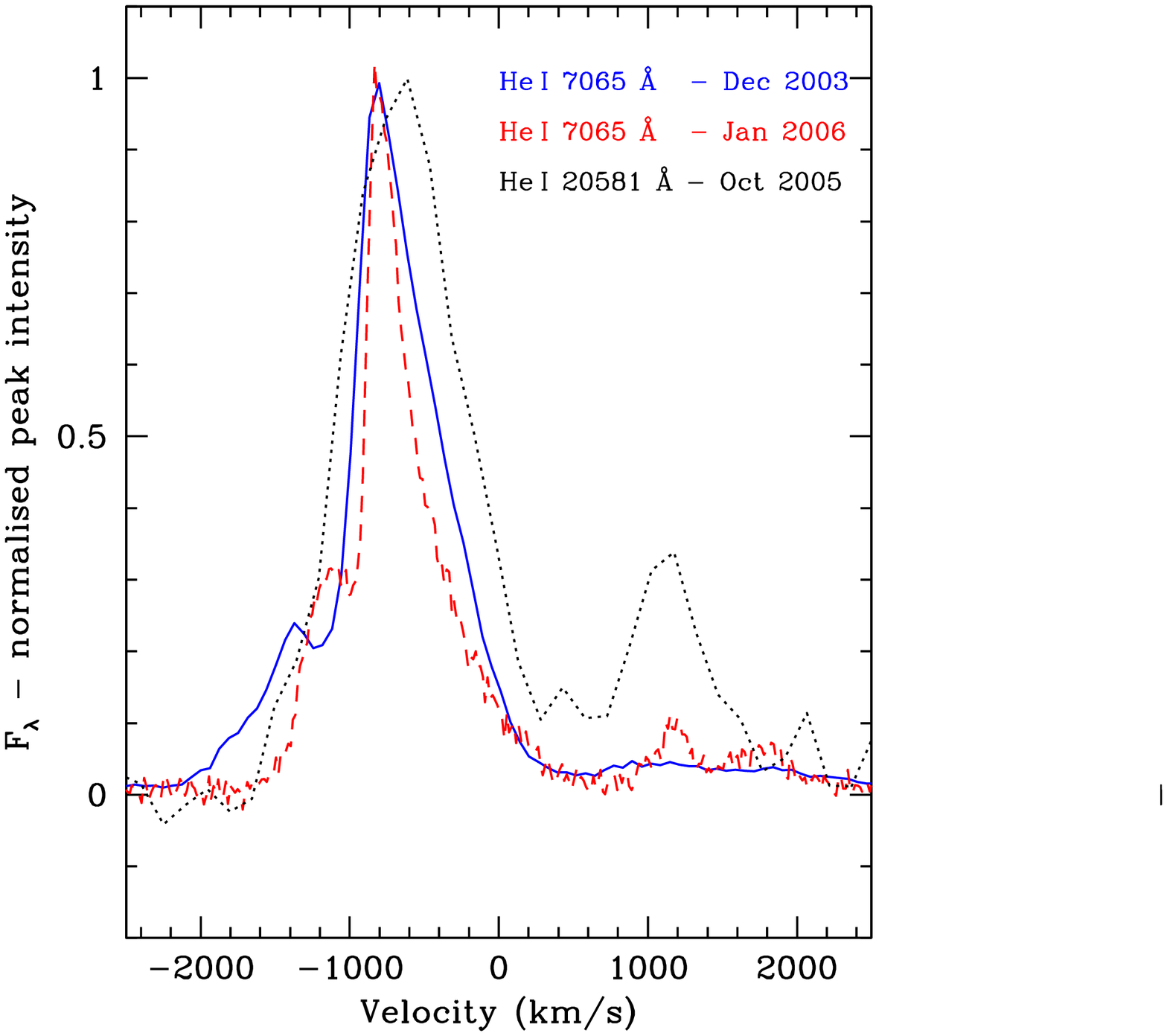}
\caption{The spatially-resolved velocity structure of the HeI 7065 {\AA} line along 
the major axis of the nova shell running from East to West 
taken in January 2006 (left panel). The right panel shows the long-slit 
line profile of HeI recombination lines in the optical (7065 {\AA}) and near-infrared (2.0581 $\mu$m) 
over a 2-year time period.}
\label{woudtfig3}
\end{figure}

\begin{figure}
\includegraphics[width=0.6\textwidth,angle=-90]{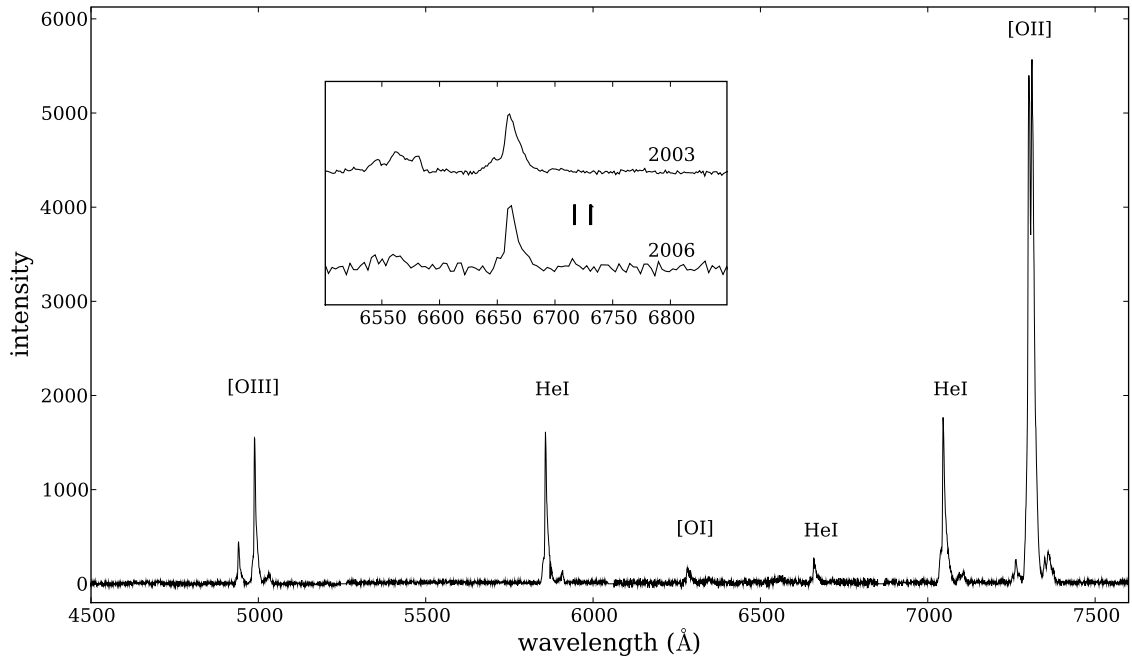}
\caption{Average spectrum of V445 Puppis obtained in January 2006 by summing the IFU fibers
containing the south-western lobe. Prominent lines are marked, including the He\,I 7065 {\AA}
line used for our kinematic analysis. The inset highlights the lack of the [S\,II] lines at
6717/6731 {\AA} (marked by the vertical bars) that would be the signature of shock emission. 
For comparison, our (deeper) December 2003 spectrum \citep{wou05} is also shown, further 
confirming the absence of [S\,II]. Note that the 2006 spectrum in the inset has been binned
to 3.1 {\AA}/pixel to improve the signal-to-noise and facilitate the comparison with the
2003 observations.}
\label{woudtfig4}
\end{figure}

In Fig.~\ref{woudtfig3} we show the spatially-resolved velocity profile of the 
He\,I 7065 {\AA} line extracted from our IFU spectroscopy. These were observed
close to the second epoch of the VLT NACO $K_s$ band imaging campaign. We averaged together
several fibers across the minor axis of the nova-shell to generate each plotted 
line profile, sampling the shell approximately along its major axis, from east to west 
on the sky. This sampling at $0.17''$ per row of fibers is seeing-limited with a FWHM 
of $0.5''$ during our observations.
Two components are distinctly visible, associated with each side of the nova
shell, as expected from a bipolar outflow \citep{solf85}.
Peak emission of the NE lobe is at $-1270$ km s$^{-1}$ and $+1140$ km s$^{-1}$,
whereas peak emission of the SW lobe is at $-760$ km s$^{-1}$ and $+1720$ km s$^{-1}$; typical
measurement errors are $\pm$ 20 km s$^{-1}$. 
The offset is presumably due to a mild inclination to the line-of-sight, which
results in a different mean radial velocity for the two lobes ($-60$ km s$^{-1}$ versus 
$+480$ km s$^{-1}$ for the NE lobe and the SW lobe, respectively). The mean of these
velocity components agrees well with the systemic radial velocity of V445 Puppis of 224 $\pm$ 8 km
s$^{-1}$, reported by \citet{iijim08}. All the emission lines in the Magellan IFU spectra 
reveal the same velocity structure, see also the summed spectrum shown in Fig.~\ref{woudtfig4}. 
The He\,I 7065 {\AA} was chosen since it is one of 
the strongest unblended lines and allows for a cleaner measurement compared to the 
strong but blended O features. Moreover, the overall structure of the emission lines 
(as deduced from a long-slit perspective) has not changed significantly over
$\sim$ 2 years (see right panel of Fig.~\ref{woudtfig3}) which suggests that the bulk outflow pattern 
has not evolved substantially.

Motivated by the observed kinematics, we model the velocity profile with a simple bipolar 
velocity field (non-variable in time)
following the description in \citet{solf85}, namely:

\begin{equation}
v_{ex} (\phi) = v_e + (v_p - v_e) \sin ^{\alpha} (|\phi|),
\end{equation}

where $\phi$ is the latitude angle (the poles are at $|\phi| = 90^{\circ}$, the
equator at $\phi = 0^{\circ}$), $v_e$ is the equatorial velocity and $v_p$ is the polar
velocity. The exponent $\alpha$ controls the degree of bipolarity: large $\alpha$ implies 
strong bipolarity.  In our models we kept $v_e$ constant at 500 km~s$^{-1}$ based on the (equatorial)
expansion velocity deduced from the P Cygni profile during outburst \citep{iijim08}. 
From the velocity profile alone (Fig.~\ref{woudtfig3}), no unique solution to the bipolar velocity field
(read: $\alpha$) exists, although each value of $\alpha$ gives a best fit value for the inclination 
$i$, $v_e$ and $v_p$. A wide range of models ($\alpha = 6 - 14$) all indicate a bipolar shell 
which is closely aligned with the plane of the sky ($i_{\rm shell} = 5.8^{\circ} - 3.7^{\circ}$, 
respectively), giving further support to the presence of an orthogonal dust structure closely 
aligned along the line of sight causing the observed extinction.

\begin{figure}
\epsscale{0.8}
\plotone{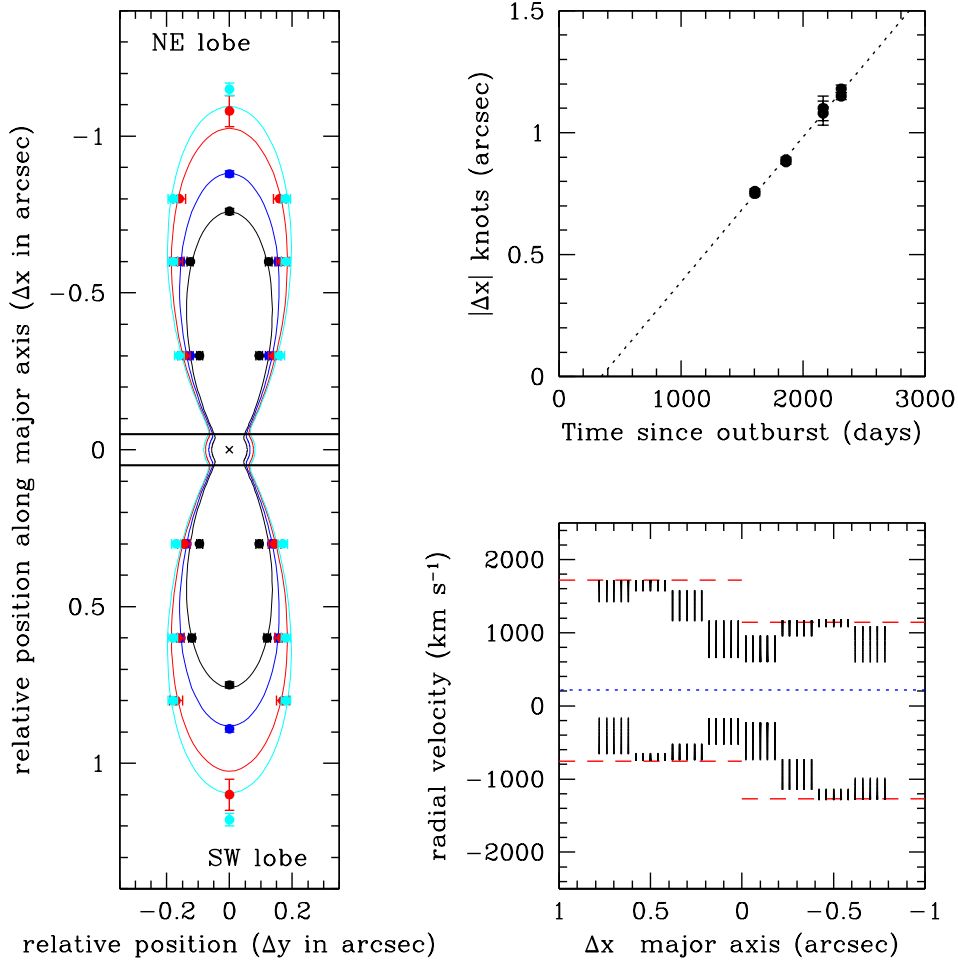}
\caption{Left panel: Measured positions of peak emission of the nova shell at selected 
distances along the major axis, including the knots at the extremes of the shell (black: March 2005,
blue: December 2005, red: October 2006, cyan: March 2007). The $\alpha = 12$ model (Table~\ref{woudttab4}) 
spatio-kinematic model is overplotted (solid lines, colors as before). The top-right panel shows the change
in position of the knots as a function of time. The lower-right panel displays the expected velocity
profile of the shell along the major axis as derived from our best fit dynamical model (Table~\ref{woudttab4}).
The horizontal dashed lines are the measured peak velocities from Fig.~\ref{woudtfig2}.}
\label{woudtfig5}
\end{figure}

Measurements of the angular separation of
peak emission perpendicular to the polar axis were obtained at $\Delta x$ = $\pm 0.3''$, $\pm 0.6''$, 
and $\pm 0.8''$ (the latter only for the last two epochs) from the center, where $x$ is the
distance along the polar (major) axis (i.e. $\phi = \pm 90^{\circ}$); they are 
listed in Table~\ref{woudttab3} and shown in the left panel of Fig.~\ref{woudtfig5}. Typical errors on the positional 
measurements are $\pm$ 0.01$''$ for the first two epochs and $\pm$ 0.02$''$ for the latter two epochs.
Models with $\alpha < 10$ are excluded from the high-resolution imaging (too round). Similarly,
models with $\alpha > 14$ are excluded as they would have resulted in such high elongation of the 
shell that it should have been visible at $\Delta x = \pm 0.8''$ at the first epoch. This was not 
observed.

From the combined spatio-kinematic observations and models, a distance to the nova shell
can be obtained; $x$ and $y$ values from the models have been derived from equations 3 and 4 from
\citet{solf85} and differences with the observed values minimised by varying the distance to the nova. 
Even though we are comparing the velocity field of {\sl optical} emission lines with the {\sl near-infrared} 
$K_s$ band images of V445 Puppis, it should be noted that the velocity structure of
the optical and near-infrared recombination lines of He\,I are identical (see 
right panel of Fig.~\ref{woudtfig3}); we are therefore probing the same outflow
component at optical and near-infrared wavelengths. The distance thus derived is relatively insensitive to the 
choice of $\alpha$; acceptable model fits ($\alpha = 10 - 14$) lead to distances
in the range of $7.8 - 8.4$ kpc. These values are consistent with the lower limit of 6 kpc deduced from
the presence of interstellar Na D absorption lines at $v_{\rm lsr} = +73.5$ km/s \citep[see,][]{brand93}
in the high-resolution spectrum of V445 Puppis during outburst \citep{iijim08}.

We superimpose the best fit model ($\alpha = 12$) on the measurements in the left panel of Fig.~\ref{woudtfig5}. The 
parameters of the $\alpha = 12$ model are given in Table~\ref{woudttab4} where the formal measurement error is
given as well as the systematic error for different allowed values of $\alpha$.  The expected radial velocity profile
from the $\alpha = 12$ model -- along the major axis of the shell -- is shown in the lower right of Fig.~\ref{woudtfig5}
and can be compared with the observed profile shown  in Fig.~\ref{woudtfig3}.
The bulk velocities ($v_p = 6720 \pm 650$ km s$^{-1}$)
are on the high end compared to hydrogen-rich novae, though not unreasonable, see for example
RS Oph with $v_p = 5600 \pm 1100$ km s$^{-1}$ \citep{bode07}.
We thus derive a distance to V445 Puppis of $8.2 \pm 0.5$ kpc (including systematic
errors) by combining our dynamical study of the emission lines from the shell with on-sky expansion 
rates measured in AO images. 

It is of interest to note that the knots behave independently from the shell. Their 
apparent linear expansion on the sky ($0.217'' \pm 0.010''$ yr$^{-1}$), shown in the upper right
panel of Fig.~\ref{woudtfig5}, results in a deprojected polar velocity at 8.2 kpc 
of $\sim$ 8450 km s$^{-1}$, larger than the 6720 km s$^{-1}$ derived from the
bipolar velocity field of the shell. Moreover, a simple extrapolation of the expanding
knots back to the center of the nova remnant coincides closely in time with a strong
radio flare observed at $\sim$ HJD 245\,2185 \citep{rup01}. This suggests the knots
originate from a $\sim$345-d delayed post-outburst outflow rather than the nuclear burning event itself. 
The bipolar shell on the other hand is consistent with uniform expansion since 
the start of the nova outburst.

Detailed 1-D hydrodynamical models following the evolution of PNe \citep{schoen05,mell04}
indicate that in the interstellar wind model of PNe the measured edges of the PN shells are due to either
ionisation or shock fronts, which move at different velocities from those measured with spectroscopy;
in such case the expansion parallax always gives a lower limit to the true distance. This was indeed
observed in the symbiotic star Hen 2-147 \citep{sant07}, where the distance derived
from the expansion parallax is a factor of 2 lower than the distance determined from the period-luminosity 
relation for its Mira star component. \citet{sant07} argue for the presence of a shock front in the
case of Hen 2-147. In the case of V445 Puppis, it is less clear that the edges in Fig.~\ref{woudtfig2} arise
from a strong shock front. Our optical spectra show a distinct lack of lines related to shock ionisation, 
e.g. both [NII] and [SII] 6717/6731 {\AA} are absent, as seen in Fig.~\ref{woudtfig4}.

A direct comparison with the \citet{schoen05} models is not appropriate given
the order of magnitude difference in outflow velocities. 
We do not - a priori - expect a large distance correction factor for V445 Puppis. The deprojected 
velocities of the polar blobs are already on the high end of speeds observed in nova ejecta; 
a significantly greater distance would imply velocities more typical of supernovae 
rather than classical novae. Moreover, the absence of strong shock lines, the consistency of the
He\,I recombination line profiles (kinematic analysis), linked with the dominance of the
He\,I recombination line in the $K_s$ passband (spatio analysis) suggests that the outflow velocity matches the 
angular expansion closely. 
Nonetheless, to estimate by what amount the expansion parallax could underestimate the true distance in V445 Puppis
detailed (magneto)hydrodynamic simulations \citep[see, e.g.,][]{denn09} with realistic input 
parameters as identified here -- and the possible presence of pre-existing circumstellar material --
must be obtained.

\section{Discussion}

\subsection{Pre-outburst conditions}

With the distance known, the nature of the helium nova progenitor can be constrained from pre-outburst
observations. Unfortunately, not much is known of V445 Puppis prior to outburst. Apart from optical ($V =
14.5$ mag, from VSNET, see \citet{ashok03}) and near-infrared (2MASS: $K_s = 11.52$ mag) 
flux measurements (see lower panel of Fig.~\ref{woudtfig1}), neither 
spectrum nor orbital period have been obtained prior to outburst. We verified the
plate archives at the Harvard-Smithsonian Center for Astrophysics for prior outbursts
(or sustained long periods of obscuration indicative of a missed outburst). Despite 
good time coverage and the fact that V445 Puppis could be identified on many plates at approximately 
constant brightness (based on visual comparision with nearby stars in the field), 
we could find no signatures of a previous outburst in the 1897 -- 1955 time frame. 

\begin{figure}
\epsscale{0.8}
\plotone{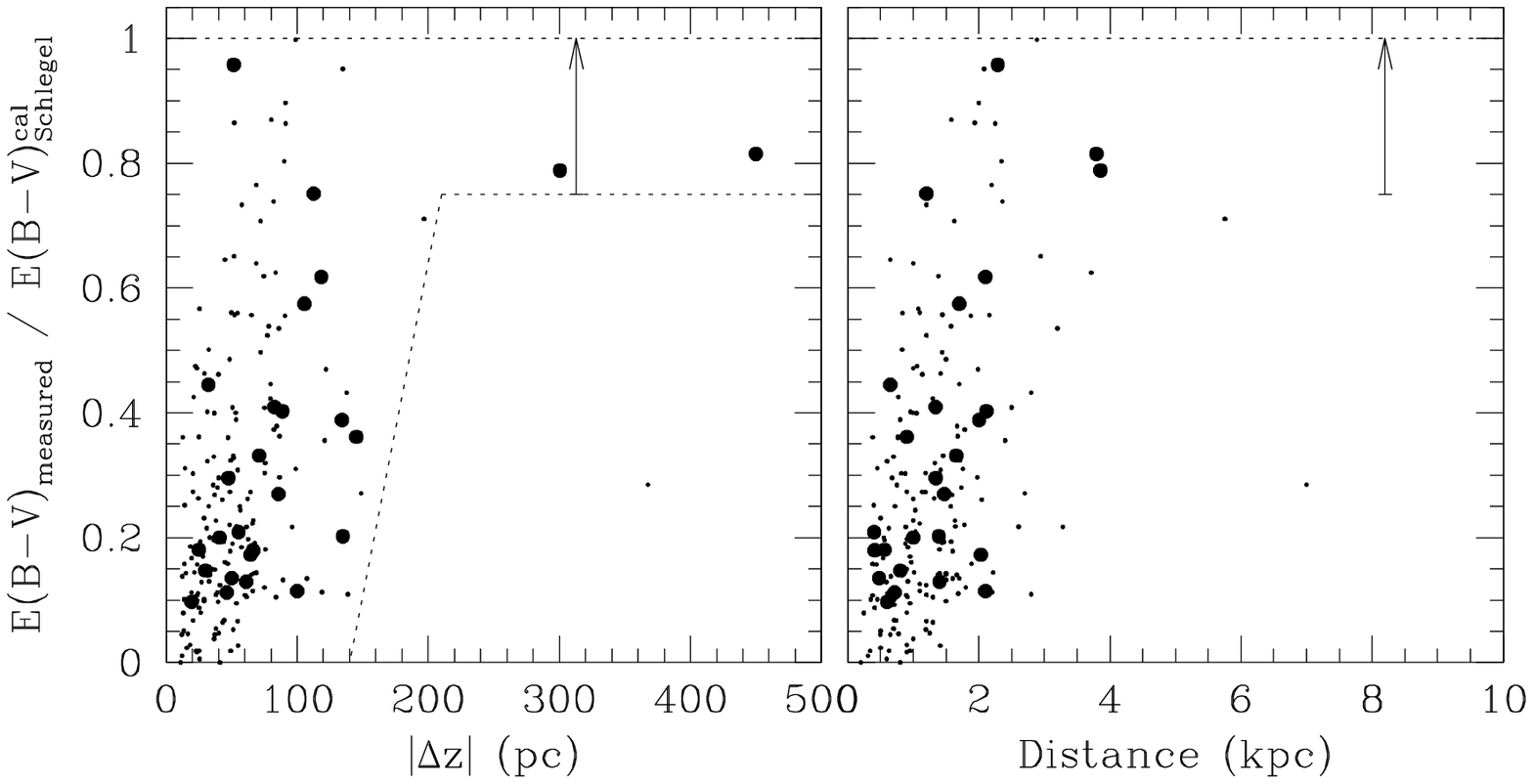}
\caption{The ratio of measured reddening to the total line-of-sight reddening for Galactic open clusters
\citep{khar05} as a function of height above/below the Galactic Plane (left panel) and distance (right panel). Only open clusters
located at $1^{\circ} \le |b| \le 4^{\circ}$ and with $E(B-V)_{\rm Schlegel}$ are shown. The 
total line-of-sight reddening has been reduced to 87\% of the \citet{schleg98} values
as recalibrated by galaxies seen through the southern Milky Way \citep{schroed07}. The big filled dots
represent open clusters within 10 degrees of V445 Puppis; the probable range of $E(B-V)$ for V445 Puppis is
indicated by the arrow.}
\label{woudtfig6}
\end{figure}

V445 Puppis is located at the low Galactic latitude of $b = -2.19^{\circ}$, which at a distance
of 8.2 $\pm$ 0.5 kpc translates to a height below the Galactic Plane of 313 $\pm$ 19 pc. At that distance,
and that far below the Galactic Plane, one can use the IRAS/DIRBE Galactic reddening maps \citep{schleg98}
to gauge the Galactic reddening towards V445 Puppis. In Fig.~\ref{woudtfig6} we show the ratio of the
measured reddening (at a given distance) to the total line-of-sight Galactic reddening for
open clusters in the Milky Way \citep{khar05} in the Galactic latitude range $1^{\circ} \le |b| \le 4^{\circ}$ and
for $E(B-V)_{\rm Schlegel} \le 2.5$, as a function of height above/below the Plane and distance, respectively.
The reddening maps of \citet{schleg98} are not well-calibrated at low Galactic latitude; from colors of 
galaxies discovered at low Galactic latitude behind the southern Milky Way it appears that the reddening values from 
Schlegel et al.~are too high \citep{schroed07} and that the true value is 87\% of the \citet{schleg98} reddening, e.g.
$E(B-V)^{\rm cal}_{\rm Schlegel} = 0.87 E(B-V)_{\rm Schlegel}$.

Comparing the location of V445 Puppis to open clusters within 10 degrees of V445 Puppis 
(big filled circles in Fig.~\ref{woudtfig6}), we deduce that Galactic foreground reddening 
towards V445 Puppis is probably in the range of $E(B-V) = 0.51 - 0.68 = 0.75 - 1.0 \, E(B-V)^{\rm cal}_{\rm Schlegel}$,
as marked by the arrow in Fig.~\ref{woudtfig6}.
The lower limit given here could be a conservative one due to a selection effect; given the relative poor
angular resolution of the reddening maps, the open clusters which set the lower ratio limits (those observed at large 
distances and large height above/below the Galactic plane) are likely to be observed through small-scale patches 
of relatively lower extinction compared to their surrounding.  Finally, taking the equivalent widths of the two interstellar
Na D components \citep{iijim08}, a value of $E(B-V) = 0.62$ is inferred using the calibration of \citet{mun97}. This value
falls within our deduced limits.

Taking the lower limit of the Galactic reddening towards V445 Puppis ($E(B-V) = 0.51$), we derive a 
pre-outburst color of V445 Puppis of $(V-K)^0 = 1.58$ mag,
assuming a standard Galactic extinction law \citep{card89}. This is significantly redder than
the likely binary progenitors -- high $\dot{M}$ AM CVn systems or systems with helium stars
donors -- which typically have $(V-K)^0 \approx -0.6$. 
A red giant donor (symbiotic nova) can be excluded on the basis of the pre-outburst colors.

This suggests the presence of substantial circumstellar (CS) reddening before the November 
2000 outburst, which is not unreasonable given the possibility of previous outbursts or 
material expelled during a common envelope phase. The CS reddening around V445 Puppis 
does not necessarily have to follow a standard Galactic extinction given the current carbon-rich
ejecta; \citet{berge99} find some deviations from a standard reddening law for CS reddening
around a number of carbon-rich R CrB stars. In the right panel of Fig.~\ref{woudtfig3}, we compare
the line profile of the 7065 {\AA} and 2.0581 $\mu$m He\,I recombination lines, obtained close in
time and both normalised by peak blueshifted emission. As expected, the redshifted component
of the optical line (at +1140 km s$^{-1}$) appears dimmer compared to its near-infrared counterpart, 
due to dust obscuration within the shell. For a standard Galactic reddening law \citep{card89}, we
expect a ratio of peak emission of near-infrared (2.0581 $\mu$m) to optical (7065 {\AA}) of 7. 
The ratio observed in V445 Puppis is 3.4, substantially less and indicative of a low ratio of
total-to-selective extinction, $R_V \approx 2.5$ \citep{fitzp99}. The shell of V445 Puppis offers an
opportunity to determine the nature of the dust extinction in carbon-rich outflows. 
The new X-shooter instrument on the VLT will be ideal to obtain simultaneous optical to near-infrared 
medium-to-high resolution spectroscopy of both NE and SW shells. 

To make the pre-outburst color of V445 Puppis consistent with the intrinsic colors
of likely progenitor binaries, a (maximum) additional color excess of $A_V - A_K$ = 2.2 mag requires $A_V$ = 2.5 mag
for $R_V = 3.1$ (standard reddening law), or $A_V$ = 2.8 mag for $R_V = 2.5$.  The uncertainty
in $A_V$ introduced by the different reddening laws, however, is small compared to the uncertainty 
in the bolometric correction needed to arrive at the pre-outburst luminosity of V445 Puppis.

We thus determine a pre-outburst extinction-corrected brightness of at least 
$V^0 = 12.9$ mag (corrected for Galactic reddening only), but more likely $V^{0}
\approx 10.1 - 10.4$ mag (including circumstellar reddening correction), with the range
in the latter allowing for differences in the CS reddening law. The corresponding absolute
magnitudes are $M_V^{0} = -1.7$ and $-4.2$ to $-4.5$ mag, respectively. 
For a range of likely bolometric corrections (BC = $-1$ to $-2.5$,
\citet{roel07,heb81}), the pre-outburst luminosity of V445 Puppis is 
$\log L/L_{\odot} = 3.3 \pm 0.3$ (corrected for Galactic
extinction only), or $\log L/L_{\odot} = 4.34 \pm 0.36$ (including circumstellar reddening correction).

It has not escaped our attention that hydrogen-deficient carbon (HDC) stars also
occupy this luminosity regime. With a larger distance -- this paper -- and a larger
Galactic reddening correction \citep{iijim08}, the possibility that V445 Puppis belongs
to the class of HDC stars can no longer be rejected on luminosity grounds, cf. \citet{ashok03}.
The (Galactic) extinction-corrected $(V-K_S)^0$ colors of V445 Puppis are not too dissimilar 
from the HDC star HD\,182040 \citep{brun98}. It is not clear, though, what can cause a luminosity
increase of 6 magnitudes and a rapidly expanding shell from a HDC star; the overall light curve
appears much better described by an event near a compact, massive white dwarf.

\subsection{Shaping of the nebula}

Deviations from the simple outflow model presented in Section 3 could provide
clues to the mechanism responsible for shaping the nebula. Although the
dynamical model fits the shell remarkably well, the largest deviations occur 
nearest to the nova remnant at the earliest
epochs. The nebula initially had a very narrow waist, followed by a rapid
broadening of the waist close to the nova remnant.

To produce very narrow waists in PNe, collimated fast winds (CFW) are needed in
systems with high density gas in an equatorial plane close to the source of the CFW
\citep{sok00}. It is unclear what collimated the fast wind in V445 Puppis. 
At the moment it seems likely that an equatorial disk/torus already existed
pre-outburst, given the large pre-outburst circumstellar reddening deduced in Section 4.1. Once the
present optically-thick dust disk clears (Fig.~\ref{woudtfig2} -- presumably additional
material was ejected in the equatorial plane during the November 2000 outburst)
and a relatively unobscured view of the nova remnant is possible, 
alternative origins of the collimated outflow could be investigated, 
e.g. the presence of a rapidly rotating magnetic white dwarf. 

The shell of V445 Puppis is unique for a classical nova. Although many novae
show bipolar shells, for example the classical nova HR Del \citep{har03} and recurrent nova
RS Oph \citep{bode07,sok08}, the shell in V445 Puppis reflects the most collimated outflow 
seen in any nova. The knots are moving at jet-like velocities of $\ga$ 8000 km s$^{-1}$.  
They can be a consequence of a jet-activity about 350 days after the main outburst 
which is not long after the sudden drop off in the $V$ band is seen \citep{ashok03} 
and coincides with the radio flare \citep{rup01}.
In the symbiotic nova CH Cygni, a drop in $V$ band magnitude is associated with the
onset of a radio jet ejection, see e.g. \citet{karov07}.

\subsection{V445 Puppis as a proto-type helium nova}

At a distance of 8.2 kpc, the maximum brightness at outburst ($V = 8.6$ mag) is consistent with
the Eddington luminosity of a massive white dwarf. Moreover, the pre-outburst luminosity 
of V445 Puppis of $\log L/L_{\odot} = 4.34 \pm 0.36$ appears to rule out 
an AM CVn progenitor and may require both a luminous accretion flow as well 
as a bright donor. For comparison, a 1.3-M$_{\odot}$ white dwarf accreting 
at $\dot{M} \sim 10^{-6}$ M$_{\odot}$ yr$^{-1}$ would give an expected accretion luminosity 
of $\log L/L_{\odot} = 3.7$. Similarly, the luminosity of a moderately massive, evolved 
helium star is around $\log L/L_{\odot} \approx 3.7$ \citep{yoon03}. 

With the distance determined in this paper, a revised mass of the dust shell 
of $1.5 \times 10^{-5}$ M$_{\odot}$ is derived following \citet{lynch04}. This is consistent
with \citet{kato03}, but we note that the mass depends strongly on the assumed
temperature of the shell and will ultimately be best constrained by mid- to far-infrared
data of V445 Puppis recently obtained by {\it Spitzer}. This mass is likely to be a lower
limit given the large optical depth of the equatorial dust disk.

The various components of V445 Puppis (the dust disk, the bipolar shell) clearly show
that the immediate environment of V445 Puppis is sculpted either directly or indirectly
by (repeated) outbursts. If V445 Puppis is representative of its class (of helium novae), 
the helium counterparts to the classical hydrogen-rich novae result
in more substantial circumbinary reddening due to the carbon-rich outflow.

Validation of the white dwarf + helium star model as the appropriate binary configuration 
of V445 Puppis will come from the determination of its orbital period. Unfortunately, this 
can only be done once the optically thick dust disk surrounding the nova remnant becomes transparent. 
Given the low inclination of the bipolar outflow, the implied inclination of the equatorial 
plane which corresponds to the orbital plane of the binary is $\sim 86^{\circ}$.  We thus expect 
this to be an eclipsing binary, which would further 
facilitate our ability to constrain the component masses.

\subsection{V445 Puppis as a candidate Ia progenitor}

There are several indirect suggestions that the white dwarf in V445 Puppis is massive: the maximum
luminosity during outburst, the large velocities of the ejected blobs, 
and its pre-outburst (accretion) luminosity.

The lack of strong soft X-ray emission pre-outburst, as is expected from a supersoft 
source \citep{yoon03}
could be explained by the substantial interstellar and circumstellar absorption 
which currently totally obscures the underlying accreting binary. V445 Puppis 
has not been detected in the ROSAT all-sky survey.

Our derived luminosity is consistent with the massive white dwarf + helium 
star model of \citet{yoon03}, but see also \citet{kato08}. It is likely that V445 Puppis
belongs to a class of binaries that, in principle, could lead towards supernovae of type Ia. 
Given that the expected lifetime of a massive white dwarf + helium star binary is short, 
these progenitors are associated with the relatively young ($\sim 10^8$ yr) population of Ia SNe. 

Whether V445 Puppis itself eventually leads to a supernova Ia event, or if the current helium nova
has pre-empted that pathway, depends critically on the mass accumulation efficiency
on the surface of the white dwarf and the mass ejected during this nova outburst. This 
remains a topic of debate.

\acknowledgments

We kindly acknowledge Tetsuya Nagata for scheduling the IRSF observations of
V445 Puppis and thank all the service observers for their observations. 
The VLT observation of V445 Puppis was also performed in service mode and we 
kindly thank the service observers for their efforts in obtaining these data. 
PAW and BW acknowledge the National Research Foundation and the University 
of Cape Town for financial support. DS acknowledges Smithsonian Astrophysical
Observatory Clay fellowship and a STFC Advanced Fellowship. MK is a member of 
the Chandra X-ray Center, which is operated by the Smithsonian
Astrophysical Observatory under NASA Contract NAS8-03060. We thank the Harvard
College Observatory plate curator for access to the plate stacks.
We thank the referee for the useful comments given.

{\it Facilities:} \facility{VLT (NAOS/CONICA): ESO Program number 
075.D-0425.A/B, 078.D-0666.A/B}, \facility{IRSF (SIRIUS)}, 
\facility{Magellan (IMACS-IFU)}.

\clearpage

\clearpage

\clearpage

\clearpage

\clearpage

\clearpage

\clearpage

\clearpage

\begin{table}
\caption{Observing log of the NAOS/CONICA observations}
\label{woudttab1}
\begin{tabular}{clccc}
\hline
Epoch & Date  & MJD (start) & FWHM$_{\rm AO}$ & $t_{int}$ (= N $\times$ NDIT $\times$ DIT) \\
      &       &             & ($''$)         &   (s)  \\
\hline
1     & 26 March 2005      & 245\,3455.066     &  0.10     & 825 (= $5 \times 11 \times 15$)   \\ 
2     & 07 December 2005   & 245\,3711.278     &  0.10     & 600 (= $10 \times 1 \times 60$)   \\
3     & 06 October 2006    & 245\,4014.356     &  0.14     & 1620 (= $9 \times 4 \times 45$)    \\
4     & 03 March 2007      & 245\,4162.015     &  0.09     & 1620 (= $9 \times 4 \times 45$)   \\
\hline 
\end{tabular}
\end{table}

\vspace{2cm}

\begin{table}
\caption{Observing log of the IRSF/SIRIUS observations}
\label{woudttab2}
\begin{tabular}{lcccccc}
\hline
Date              & HJD (start)   &   $t_{int}$  &  FWHM &  $J$              &     $H$            &       $K_s$         \\
                  &               &     (s)     & ($''$) & (mag)             &   (mag)            &        (mag)        \\
\hline
22 March 2002     & 245\,2356.266 &   1000      &  0.97  &         --         & 17.419 $\pm$ 0.025 & 13.859 $\pm$ 0.027  \\
19 June 2002      & 245\,2445.185 &    600      &  1.31  &         --         & 17.907 $\pm$ 0.078 & 15.168 $\pm$ 0.025  \\
28 August 2002    & 245\,2515.657 &    900      &  1.38  & 20.650 $\pm$ 0.272 & 17.691 $\pm$ 0.045 & 14.957 $\pm$ 0.027  \\
04 November 2002  & 245\,2583.610 &    900      &  1.10  & 19.394 $\pm$ 0.084 & 16.205 $\pm$ 0.016 & 13.327 $\pm$ 0.030  \\
25 January 2003   & 245\,2665.326 &    900      &  1.51  & 20.116 $\pm$ 0.235 & 16.223 $\pm$ 0.016 & 13.021 $\pm$ 0.022  \\
13 March 2003     & 245\,2712.289 &    900      &  1.47  & 19.709 $\pm$ 0.136 & 15.907 $\pm$ 0.011 & 12.891 $\pm$ 0.031  \\
14 May 2003       & 245\,2774.242 &    900      &  1.08  & 18.834 $\pm$ 0.067 & 15.529 $\pm$ 0.014 & 12.611 $\pm$ 0.033  \\
12 November 2003  & 245\,2956.479 &   1200      &  2.02  & 19.052 $\pm$ 0.123 & 15.672 $\pm$ 0.015 & 12.715 $\pm$ 0.023  \\
25 January 2004   & 245\,3030.336 &   1200      &  1.80  & 18.825 $\pm$ 0.060 & 15.914 $\pm$ 0.010 & 13.017 $\pm$ 0.026  \\
26 April 2004     & 245\,3122.228 &    900      &  1.38  & 18.724 $\pm$ 0.103 & 15.788 $\pm$ 0.026 & 12.816 $\pm$ 0.025  \\
02 May 2004       & 245\,3128.205 &    900      &  1.76  & 18.740 $\pm$ 0.126 & 15.786 $\pm$ 0.027 & 12.822 $\pm$ 0.034  \\
29 September 2004 & 245\,3277.643 &    900      &  1.16  & 18.341 $\pm$ 0.046 & 15.580 $\pm$ 0.010 & 12.730 $\pm$ 0.028  \\
31 October 2004   & 245\,3310.582 &    900      &  1.87  & 18.291 $\pm$ 0.050 & 15.800 $\pm$ 0.013 & 12.973 $\pm$ 0.025  \\
14 November 2004  & 245\,3324.443 &    900      &  1.13  & 18.024 $\pm$ 0.039 & 15.745 $\pm$ 0.013 & 12.976 $\pm$ 0.016  \\
25 December 2004  & 245\,3365.395 &   1200      &  1.37  & 18.001 $\pm$ 0.041 & 15.675 $\pm$ 0.009 & 12.974 $\pm$ 0.025  \\
17 January 2005   & 245\,3388.458 &    900      &  1.37  & 17.923 $\pm$ 0.036 & 15.656 $\pm$ 0.008 & 12.914 $\pm$ 0.018  \\
25 February 2005  & 245\,3426.799 &    900      &  1.14  & 17.870 $\pm$ 0.034 & 15.569 $\pm$ 0.011 & 12.804 $\pm$ 0.026  \\
22 March 2005     & 245\,3451.865 &    900      &  1.10  & 17.829 $\pm$ 0.036 & 15.516 $\pm$ 0.009 & 12.837 $\pm$ 0.019  \\
14 April 2005     & 245\,3473.789 &   1200      &  1.14  & 17.742 $\pm$ 0.031 & 15.432 $\pm$ 0.007 & 12.775 $\pm$ 0.021  \\
06 June 2005      & 245\,3527.679 &    900      &  1.13  & 17.659 $\pm$ 0.035 & 15.361 $\pm$ 0.007 & 12.796 $\pm$ 0.026  \\
30 September 2005 & 245\,3644.622 &    900      &  1.17  & 17.648 $\pm$ 0.035 & 15.312 $\pm$ 0.012 & 12.826 $\pm$ 0.027  \\
25 October 2005   & 245\,3669.579 &    900      &  1.98  & 17.731 $\pm$ 0.040 & 15.323 $\pm$ 0.014 & 12.850 $\pm$ 0.031  \\
20 March 2006     & 245\,3814.745 &    900      &  1.19  & 17.686 $\pm$ 0.037 & 15.432 $\pm$ 0.011 & 12.959 $\pm$ 0.022  \\
06 October 2006   & 245\,4015.117 &    900      &  1.39  & 17.860 $\pm$ 0.034 & 15.662 $\pm$ 0.013 & 13.145 $\pm$ 0.024  \\
13 March 2007     & 245\,4172.804 &    900      &  1.42  & 17.995 $\pm$ 0.039 & 15.835 $\pm$ 0.010 & 13.323 $\pm$ 0.020  \\
16 September 2007 & 245\,4360.152 &    900      &  0.98  & 17.981 $\pm$ 0.036 & 15.975 $\pm$ 0.014 & 13.535 $\pm$ 0.030  \\
19 March 2008     & 245\,4544.781 &    900      &  1.16  & 18.000 $\pm$ 0.036 & 16.150 $\pm$ 0.011 & 13.745 $\pm$ 0.022  \\
17 September 2008 & 245\,4727.150 &    900      &  1.44  & 17.998 $\pm$ 0.041 & 16.311 $\pm$ 0.018 & 13.948 $\pm$ 0.023  \\
11 February 2009  & 245\,4873.778 &    900      &  2.03  & 17.969 $\pm$ 0.045 & 16.525 $\pm$ 0.020 & 14.128 $\pm$ 0.018  \\
\hline 
\end{tabular}
\end{table}

\clearpage

\begin{table}
\caption{Dimensions of the nova shell of V445 Puppis}
\label{woudttab3}
\begin{tabular}{cccccccccc}
\hline
Epoch &  \multicolumn{2}{c}{$\Delta$y at $|$x$|$ = 0.3$''$} & \multicolumn{2}{c}{$\Delta$y at $|$x$|$ = 0.6$''$}  &
 \multicolumn{2}{c}{$\Delta$y at $|$x$|$ = 0.8$''$} &  \multicolumn{2}{c}{$\Delta$x knots}  & errors  \\
       & ($''$)  & ($''$)  & ($''$)  & ($''$)  & ($''$)  & ($''$)  & ($''$)  & ($''$) & ($''$) \\
\hline
      &   NE    &    SW     &   NE     &    SW    &  NE   & SW  &  NE  &  SW  & \\
\hline
1  &   0.19      &   0.19        &   0.25       &   0.24       &   --  &  --   &  0.76      &  0.75   & 0.01  \\
2  &   0.25      &   0.27        &   0.31       &   0.31       &   --  &  --   &  0.88      &  0.89   & 0.01  \\
3  &   0.29      &   0.29        &   0.34       &   0.33       &  0.32     &  0.34     &  1.08$^\star$      &  1.10$^{\star}$  & 0.02   \\
4  &   0.32      &   0.34        &   0.36       &   0.36       &  0.36     &  0.36     &  1.15      &  1.18  & 0.02   \\
\hline 
\end{tabular}
\newline
{\footnotesize $^{\star}$ The uncertainties in the positions of the knots at this epoch are $\pm$ 0.05$''$.}
\end{table}

\clearpage

\begin{table}
\caption{Bipolar outflow model and properties of V445 Puppis}
\label{woudttab4}
\begin{tabular}{cc}
\hline
PA shell         & 66 deg  \\[3pt]
$\alpha$         & 12 $|^{+2}_{-2}$ \\[3pt]
$i_{\rm shell}$    & 3.9 $|^{-0.2}_{+0.5}$ $\pm$ 0.4 deg\\[3pt]
$v_e$ & 500 km s$^{-1}$ \\[3pt]
$v_p$ & 6720 $|^{+540}_{-600}$ $\pm$ 250 km s$^{-1}$  \\[3pt]
$v_{\rm knots}$ & 8450 $|^{+210}_{-410}$ $\pm$ 390 km s$^{-1}$    \\[3pt]
$d$   & 8.2 $|^{+0.2}_{-0.4}$ $\pm$ 0.3 kpc \\[3pt]
\hline
\end{tabular}
\end{table}

\end{document}